\documentclass[a4paper,twocolumn,showpacs,amsmath,pra,amssymb]{revtex4-1} 
\usepackage[T1]{fontenc}
\usepackage[latin9]{inputenc} 
\usepackage{times}
\usepackage{color}
\usepackage{xspace}
\usepackage{color,ulem}
\usepackage{amssymb,amsmath}
\usepackage{amsbsy}
\usepackage{graphicx}
\usepackage{bm}
\usepackage{epstopdf}
\usepackage{float}

\newcommand{\diff}{\mathrm{d}}

\newcommand{\bx}{\ensuremath{\mathbf{x}}\xspace}

\newcommand{\bk}{\mathbf{k}}

\begin{document}
\newcommand{\nt}{\tilde{n}}
\newcommand{\Lnh}{\ensuremath{\mathcal{L}}\xspace}
\newcommand{\dbx}{\diff\bx}
\newcommand{\dbk}{\diff\bk}
\newcommand{\nbe}{\ensuremath{\bar{n}_{\mathrm{BE}}}\xspace}

\newcommand{\bro}{\mathbf{x}_1}
\newcommand{\brt}{\mathbf{x}_2}

\title{Depletion and fluctuations of a trapped dipolar Bose-Einstein condensate in the roton regime} 
\author{P.~B.~Blakie}  
\email{blair.blakie@otago.ac.nz}
\author{D.~Baillie}     
\author{R.~N.~Bisset}  

\affiliation{Jack Dodd Centre for Quantum Technology, Department of Physics, University of Otago, Dunedin, New Zealand.}

\begin{abstract}
We consider the non-condensate density and density fluctuations of a trapped dipolar Bose-Einstein condensate, focusing on the regime where a roton-like excitation spectrum develops. Our results show that a characteristic peak in the non-condensate density occurs at trap center due to the rotons. In this regime we also find that the anomalous density becomes positive and peaked, giving rise to enhanced density fluctuations. We  calculate the non-condensate density in momentum space and show that a small \textit{momentum halo} is associated with the roton excitations.
\end{abstract}

\maketitle

\section{Introduction}
 Bose-Einstein condensates (BECs) have been produced with highly magnetic atoms  \cite{Griesmaier2005a,*Bismut2010a,Mingwu2011a,*Lu2012a,Aikawa2012a} in which the constituent  particles interact with a long ranged and anisotropic dipole-dipole interaction (DDI). By virtue of this interaction, these systems are expected to open up a rich array of new physics to explore  using ultra-cold gases \cite{Lahaye_RepProgPhys_2009}. In particular  an important prediction, which has yet to be verified in experiments, is that a rotonic-like excitation spectrum will develop when a condensate is tightly confined along the direction that the dipoles are polarized \cite{Santos2003a}.   There has been a number of recent theoretical studies proposing schemes to detect rotons and characterize their effects on the properties of dipolar BECs (e.g.~see \cite{Ronen2007a,Lu2010a,Wilson2012a,Martin2012a,Wilson2010a,Ticknor2011a,Wilson2008a,Nath2010a,Blakie2012a,Bisset2013a}). Experimentally, rotons will likely be realized in pancake shaped traps (or an array of pancake traps, e.g.~see \cite{Muller2011a}), and the trap itself plays a very fundamental role in the nature of the rotons that emerge \cite{Ronen2007a}.  In particular the rotons are sensitive to the condensate density, which decreases away from trap center due to the weak radial trapping.
As a result  the rotons are effectively confined  to the central region of the condensate, realizing what has been termed a confined \textit{roton gas} \cite{JonaLasinio2013}.  
 
In this paper we consider a pancake dipolar BEC in the regime where rotons develop. We report on the behavior of a number of key quantities that characterize the system, including the non-condensate density and density fluctuations. We show that the non-condensate density has a prominent  peak at the trap center when the excitation spectrum has a roton feature, which we refer to as the \textit{roton peak}. We verify that this peak is due to the excitations in the neighborhood of the roton minimum.
We also  find that the anomalous density, which is normally negative for repulsive interactions, turns positive in the roton regime, signaling enhanced density fluctuations. Indeed, we show that the relative size of the density fluctuations can  be a  significant fraction of the total density even at low temperatures, where the condensate depletion is still small.  Our fully three-dimensional (3D) calculations are performed within the framework of a Bogoliubov treatment of the excitations, and our results show that the rotons have an essential 3D character. Our work will hence provide a useful benchmark for calculations made within the quasi-two-dimensional (quasi-2D) approximation, where motion is assumed frozen out in the tightly confined direction (also see \cite{Fischer2006a,Wilson2011a}).
We note that other work in this area has focused on close-to-spherical traps \cite{Ronen2007b,Lima2011a,Lima2012a} or has been in the quasi-2D regime \cite{Ticknor2012a,Ticknor2012b,Sykes2012a}.

\section{Meanfield Formalism}  
Here we briefly review the meanfield theory of a dipolar BEC and refer to Ref.~\cite{Ronen2006a} for a more detailed discussion.
Within a meanfield approach, valid for weakly interacting condensates, the field operator takes the form  
 \begin{equation}
 \hat\Psi(\mathbf{x}) \approx \psi_0(\mathbf{x}) + \hat{\delta}(\mathbf{x}).\label{psihatbog}
\end{equation}

\subsection{Gross-Pitaevskii equation}  
In Eq.~(\ref{psihatbog}) $\psi_0$ is the condensate wavefunction (normalized to the condensate number $N_0$), which satisfies the non-local dipolar Gross-Pitaevskii equation (GPE) \cite{Goral2000a}
\begin{align}
 \mu \psi_0&=\left[\!-\frac{\hbar^2\nabla^2}{2m}\!+\!V(\mathbf{x})\!+\!\int d\mathbf{x^\prime} U(\mathbf{x}\!-\!\mathbf{x^\prime})| \psi_0(\mathbf{x^\prime})|^2\right]\psi_0 ,\\
 &\equiv \mathcal{L}_{\mathrm{GP}}\psi_0,
\end{align}
where  $\mu$ is the chemical potential.  For dipoles  polarized along $z$  the inter-atomic interaction potential is of the form $U(\mathbf{r}) =g\delta(\mathbf{r})+U_{\mathrm{dd}}(\mathbf{r})$,
where the short range interaction is characterized by the contact parameter $g=4\pi a_s\hbar^2/m$, with $a_s$ being the $s$-wave scattering length. The DDI potential is
\begin{equation} 
U_{\mathrm{dd}}(\mathbf{r}) = \frac{3g_{\rm dd}}{4\pi}\frac{1-3\cos^2\theta}{|\mathbf{r}|^3},
\end{equation} 
where  $g_{\rm dd} = \mu_0\mu_m^2/3$, with $\mu_m$ being the magnetic dipole moment, and $\theta$ is the angle between $\mathbf{r}$ and the $z$ axis.  
The atoms are taken to be harmonically confined by the potential 
\begin{equation}
V(\mathbf{\mathbf{x}}) = \frac{1}{2}m\omega_\rho^2(\rho^2+\lambda^2z^2), 
\end{equation}
with  $\lambda=\omega_z/\omega_\rho$ being the aspect ratio.

\subsection{Bogoliubov quasiparticles}  
The fluctuation operator is expanded as
\begin{equation}
\hat{\delta}(\mathbf{x})=\sum_i\left[u_i(\mathbf{x})\hat\alpha_i -v_i^*(\mathbf{x})\hat\alpha_i^\dagger\right],
\end{equation}
where $\{u_j,v_j\}$ are the quasiparticle modes, with respective energies $\epsilon_j$, and $\hat{\alpha}_j$ are the bosonic quasiparticle mode operators. These quasiparticles are the elementary  linearized excitations of the condensate and are obtained by solving the Bogoliubov-de Gennes (BdG) equations 
\begin{align}
\mathcal{L}\left(\begin{array}{c}u_i\\ v_i\end{array}\right)&=\epsilon_i\left(\begin{array}{c}u_i\\ v_i\end{array}\right),\label{Eq_BdG}
\end{align}
where
\begin{equation} \mathcal{L}=\left(\begin{array}{cc}
\mathcal{L}_{\mathrm{GP}}-\mu+X& -X\\
X &-(\mathcal{L}_{\mathrm{GP}}-\mu+X)
 \end{array} \right),\label{Eq_LBdG}\end{equation}
 with the exchange operator $X$   defined by
\begin{equation}
Xf \equiv  \psi_0(\mathbf{x})\int d\mathbf{x}^\prime U(\mathbf{x}-\mathbf{x^\prime}) f(\mathbf{x}^\prime) \psi_0^*(\mathbf{x^\prime}).
\end{equation} 
To quadratic order the quasiparticle basis diagonalizes the  many-body Hamiltonian with $\langle \hat{\alpha}_i^\dagger\hat{\alpha}_j\rangle=\delta_{ij}\bar{n}_j$ in thermal equilibrium, where $\bar{n}_j =\left(e^{\beta\epsilon_j}-1\right)^{-1}$ and $\beta=1/k_BT$  is the inverse temperature.

\subsection{Parameter Regime and Numerical methods}  
In the results that follow we adopt harmonic oscillator units defined by the radial trap frequency, in particular $a_{\rho}=\sqrt{\hbar/m\omega_{\rho}}$ as the unit of length. We follow Ref.~\cite{Ronen2006a} and introduce $C=Na_s/a_\rho$ and $D=3Ng_{\rm dd}m/4\pi\hbar^2a_\rho$ as the dimensionless contact interaction and DDI parameters, respectively.

In this paper we focus on the case $\lambda=20$ and values of $D\le320$. We can put these parameters  into the context of current experiments: The important case we consider of $D=220$ would correspond to about $25\times10^3$ $^{164}$Dy atoms in a trap with $\omega_{\rho}=2\pi\times11$ s$^{-1}$. For this case the maximum temperature we consider of $T=10 \hbar\omega_{\rho}/k_B$ corresponds to $5.3$ nK, which is about 14\% of the condensation temperature $T_c$. We have checked that qualitatively similar results to those presented in this paper are obtained  for $\lambda=40$, and we expect our results to capture the behavior of weakly interacting dipolar condensates with  $\lambda\gtrsim 10$ (also see \cite{Petrov2000a,Fischer2006a}).

Our numerical algorithm closely follows the cylindrical method presented by Ronen \textit{et al.}~\cite{Ronen2006a}. We employ the cylindrically cutoff DDI, as described in \cite{Lu2010a}, to improve the accuracy of our numerics in the pancake geometry. We also ensure our quasi-particles are orthogonal to the condensate (see \cite{Martin2012a}). It is important to have suitably dense spatial grids to ensure that the short wavelength rotons are well represented, and we use cylindrical grids with 650 points radially over the range $\rho/a_{\rho}=[0,26]$, and 50 points along the positive $z$ axis with range $z/a_{\rho}=[0,2.46]$. The results we present here are typically calculated using the lowest $\sim30\times10^3$ quasi-particle modes, including all modes with energies up to $120\hbar\omega_{\rho}$.

\section{Results}
\subsection{Non-condensate density}
 \begin{figure*}[!tbh]
\begin{center}
\includegraphics[width=5.6in]{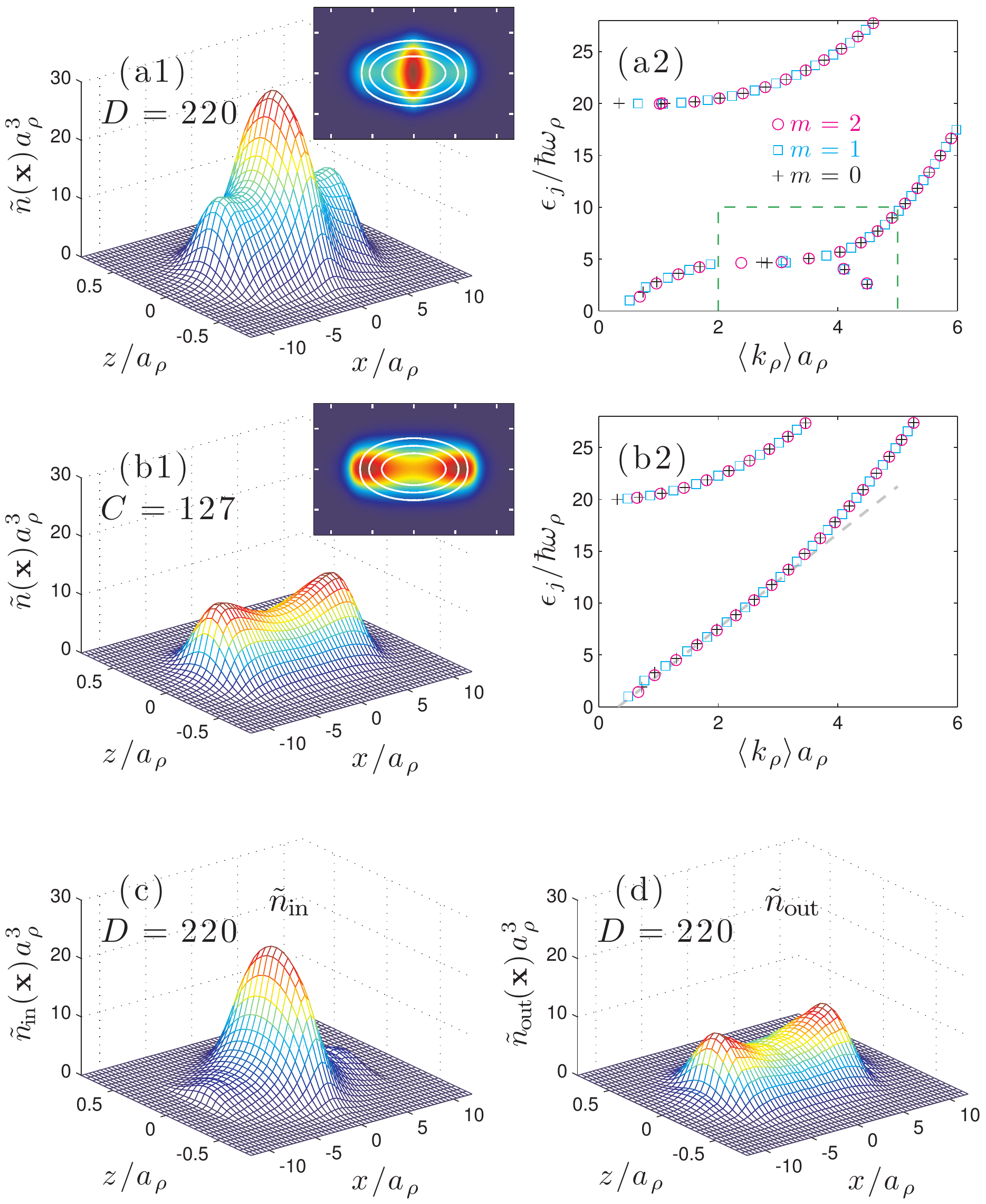}
\caption{(color online) The non-condensate density in the $y=0$ plane  for a system with (a1) dipolar interactions   $D=220$ ($C=0$) and (b1) contact interactions $C=127$ ($D=0$). 
 Insets show the the same data in the main subplot as a false color image with white contours added to show the condensate density in each case [also see Fig.~\ref{Fig:posden2}(a) and (b) for the condensate density along the $x$ axis for (a1) and (b1), respectively]. 
The spectrum of excitations mapped to a dispersion for the (a2) dipolar case and (b2) contact case, only showing modes with angular momentum projection $|m|\le2$.  The dashed line in (b2) is a visual guide to indicate that the excitation energies increase approximately linearly with radial momentum. (c)  $\tilde{n}_{\mathrm{in}}$ and (d) $\tilde{n}_{\mathrm{out}}$ (see text) for the case shown in (a1), with the energy and $k_{\mathrm{\rho}}$ range of modes used to construct these shown by the dashed box in (a2).
Other parameters $\lambda=20$ and $T=10\,\hbar\omega_{\rho}/k_B$.  
 \label{Fig:posden}}
\end{center}
\end{figure*}
 
The non-condensate density, given by
\begin{equation}
\tilde{n}(\mathbf{x})\equiv\langle\hat{\delta}^\dagger\hat{\delta}\rangle=\sum_j\left[\bar{n}_j|u_j(\mathbf{x})|^2+(\bar{n}_j\!+\!1)|v_j(\mathbf{x})|^2\right],\label{ntilde}
\end{equation}
characterizes the atoms excited out of the condensate due to interactions and thermal effects.
In Figs.~\ref{Fig:posden}(a1) and (b1) we show the  non-condensate density for systems in a pancake shaped trap at a small, but non-zero temperature. The result in Fig.~\ref{Fig:posden}(a1) is for a condensate with only dipole interactions (i.e.~$D=220$, $C=0$), while Fig.~\ref{Fig:posden}(b1) is for a contact interaction only case (i.e.~$D=0$, $C=127$). 
The values of the interaction parameters were chosen to ensure than both cases had approximately the same chemical potential ($\mu\approx37.5\hbar\omega_{\rho}$),  which leads to the condensate modes being quite similar [e.g.~see condensate density contours shown in the insets to Figs.~\ref{Fig:posden}(a1) and (b1)]. 
The value $D=220$ is sufficiently large to reveal the effects of DDIs on the system, yet is still well in the stable region (about 10\% below the critical value  at which the system becomes dynamically unstable \cite{Bisset2013a}). Indeed, for these parameters the system gas exhibits a roton like feature in its excitation spectrum [see Fig.~\ref{Fig:posden}(a2), and description below], and  associated with this is a prominent peak in the non-condensate density near the trap center [see Fig.~\ref{Fig:posden}(a1)], which we refer to as the \textit{roton peak}. 
This  {roton peak} in $\tilde{n}$  is absent in the contact only case, which instead has a local minimum (along the $x$-axis) at the trap center [see Fig.~\ref{Fig:posden}(b1)].
It is also important to take note that the roton peak exhibits a greater $z$-extent than the parts of $\tilde{n}$ away from the trap center. This indicates that the excitations responsible for the roton peak have some important 3D character and would not be accurately captured in the quasi-2D treatment in which the $z$-motion is essentially frozen out (also see \cite{Wilson2011a}). This reaffirms the necessity of the full 3D numerical solution we provide here.

To quantify the nature of excitations in the trapped system we follow the procedure introduced in \cite{Wilson2010a} to approximately map the excitations to a dispersion relation. In this procedure  each quasiparticle is assigned a momentum according to 
\begin{equation} 
\langle k_{\rho}\rangle_j\equiv\sqrt{\frac{\int d\mathbf{k}\,k_{\rho}^2\left[| {\bar u}_j(\mathbf{k})|^2+|\bar v_j(\mathbf{k})|^2\right]}{\int d\mathbf{k}\, \left[| \bar{u}_j(\mathbf{k})|^2+|\bar v_j(\mathbf{k})|^2\right]} }, 
\end{equation}
where $\bar{u}_j(\mathbf{k})=\mathcal{F}[u_j(\mathbf{x})]$, $\bar{v}_j(\mathbf{k})=\mathcal{F}[v_j(\mathbf{x})]$ are the quasiparticle amplitudes in momentum space, with $\mathcal{F}$ representing  the Fourier transform.
The result of this analysis for the purely dipolar case [see Fig.~\ref{Fig:posden}(a2)] reveals a clear flattening in the dispersion relation for $2/a_{\rho}\lesssim \langle k_{\rho}\rangle_j\lesssim5/a_{\rho}$, as well as some particular modes dropping down. The upper branch of excitations ($\epsilon_j\ge20\hbar\omega_\rho$) in this figure corresponds to modes that are excited in the tightly confined $z$-direction.
 For the contact interaction case [see Fig.~\ref{Fig:posden}(b2)] a strong phonon-like (linear)  dispersion is apparent, with no roton-like softening in the $\langle k_{\rho}\rangle_j$ range where it  occurs for the dipolar case.

  \begin{figure*}[!tbh]
\begin{center}
\includegraphics[width=5.6in]{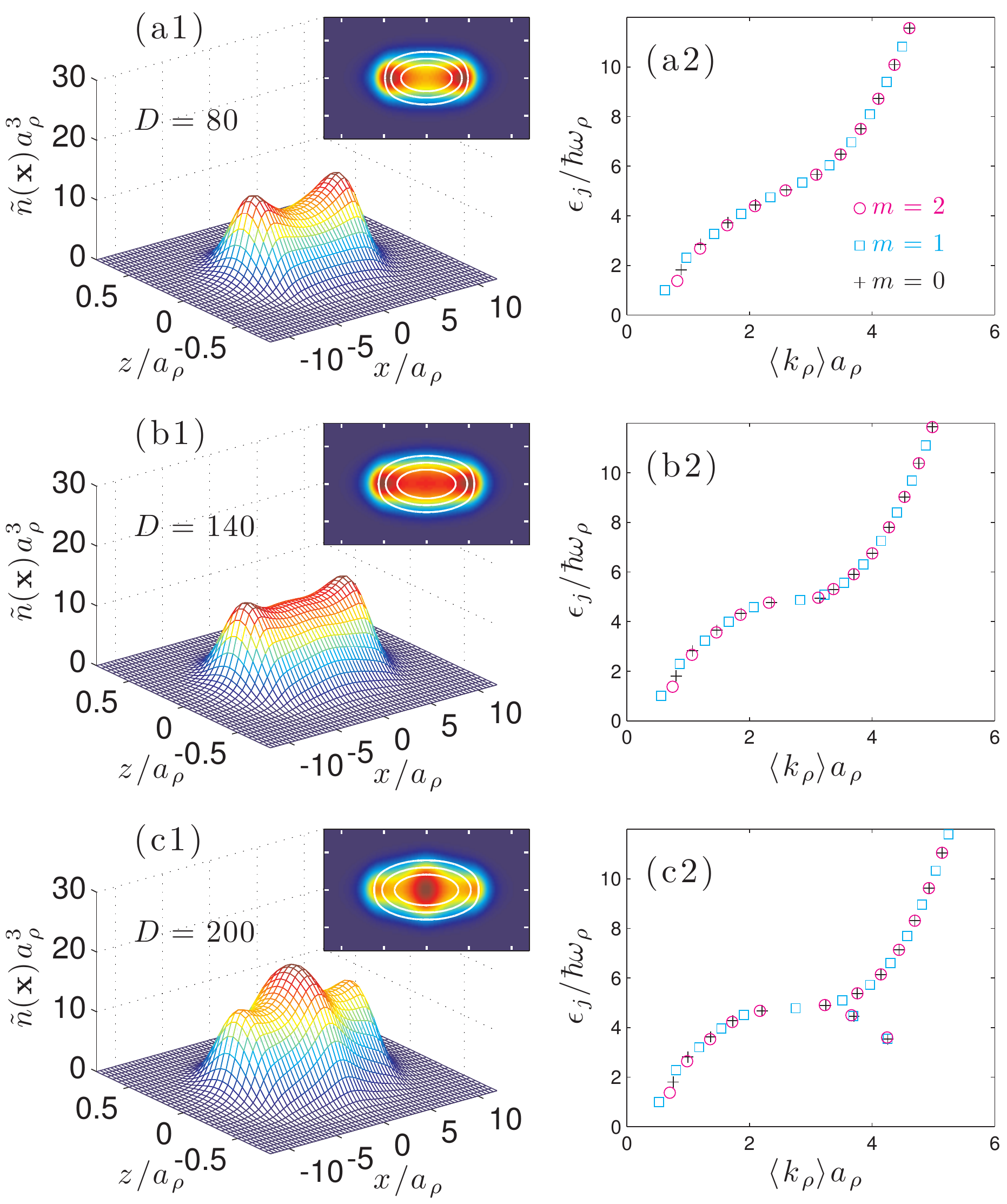} 
\caption{(color online) Development of a roton peak in the non-condensate as the dipole interaction increases for $T=10\,\hbar\omega_{\rho}/k_B$. Non-condensate density in the $y=0$ plane for (a1) $D=80$, (b1) $D=140$, (c1) $D=200$, with respective mapped dispersions in (a2)-(c2). Insets show the non-condensate density with white lines indicating contours of the condensate density. Note: $C=0$ in all these results.
 \label{Fig:func_of_D}} 
\end{center}
\end{figure*}

To verify the relationship between the roton-like part of the spectrum and the roton peak in the  non-condensate density we define $\tilde{n}_{\mathrm{in}}$ as the non-condensate density calculated according to Eq.~(\ref{ntilde}) but with the summation over modes restricted to those satisfying $\langle k_{\rho}\rangle_j\in[2/a_{\rho},5/a_{\rho}]$ and $\epsilon_j<10\hbar\omega_{\rho}$. The  boundaries of this region are indicated by the dashed box in Fig.~\ref{Fig:posden}(a2), and  were selected to include the modes where the  roton-like softening is observed. We define $\tilde{n}_{\mathrm{out}}$ to be the non-condensate density arising from all other modes, such that $\tilde{n}=\tilde{n}_{\mathrm{in}}+\tilde{n}_{\mathrm{out}}$. In Fig.~\ref{Fig:posden}(c) we show $\tilde{n}_{\mathrm{in}}$ for the dipolar case, verifying that the roton peak arises from modes in the roton region. Furthermore, $\tilde{n}_{\mathrm{out}}$, shown in Fig.~\ref{Fig:posden}(d), is very similar to the contact case [c.f.~Fig.~\ref{Fig:posden}(b1)], since the character of long wavelength  excitations  (i.e.~those with $\langle k_{\rho}\rangle_j<2/a_{\rho}$) for the two cases [compare Figs.~\ref{Fig:posden}(a2) and (b2)] are similar.

The localized nature of the roton-like excitations for a pancake shaped dipolar BEC was revealed in previous numerical studies that examined individual excitations \cite{Ronen2007a,Blakie2012a}. More recently, Jona-Lasinio \textit{et al.}~\cite{JonaLasinio2013} have proposed the idea of a confined  \textit{roton gas} based on a local density analysis of a trapped dipolar condensate. Their analysis shows that the sensitivity of the roton excitations to the condensate density effectively confines these excitations to a small region near the trap center, explaining the roton peak in $\tilde{n}$ we observe.

In Fig.~\ref{Fig:func_of_D} we show how the density peak in $\tilde{n}$ develops for several values of the dipolar interaction strength. The peak is absent for low interaction strengths, where the  excitation spectrum is monotonically increasing with  $\langle k_{\rho}\rangle$ [Fig.~\ref{Fig:func_of_D}(a1), (a2)].  The peak first appears at $D\approx140$ [see Fig.~\ref{Fig:func_of_D}(b1)]  which is also when the excitation spectrum flattens to a horizontal plateau [i.e.~at $\langle k_{\rho}\rangle\sim2.5/a_{\rho}$ in Fig.~\ref{Fig:func_of_D}(b2)] and becomes more prominent at larger interaction strengths  [Fig.~\ref{Fig:func_of_D}(c1), (c2)].

\subsection{Anomalous density and fluctuations}
\begin{figure}[!tbh]
\begin{center}
\includegraphics[width=3.45in]{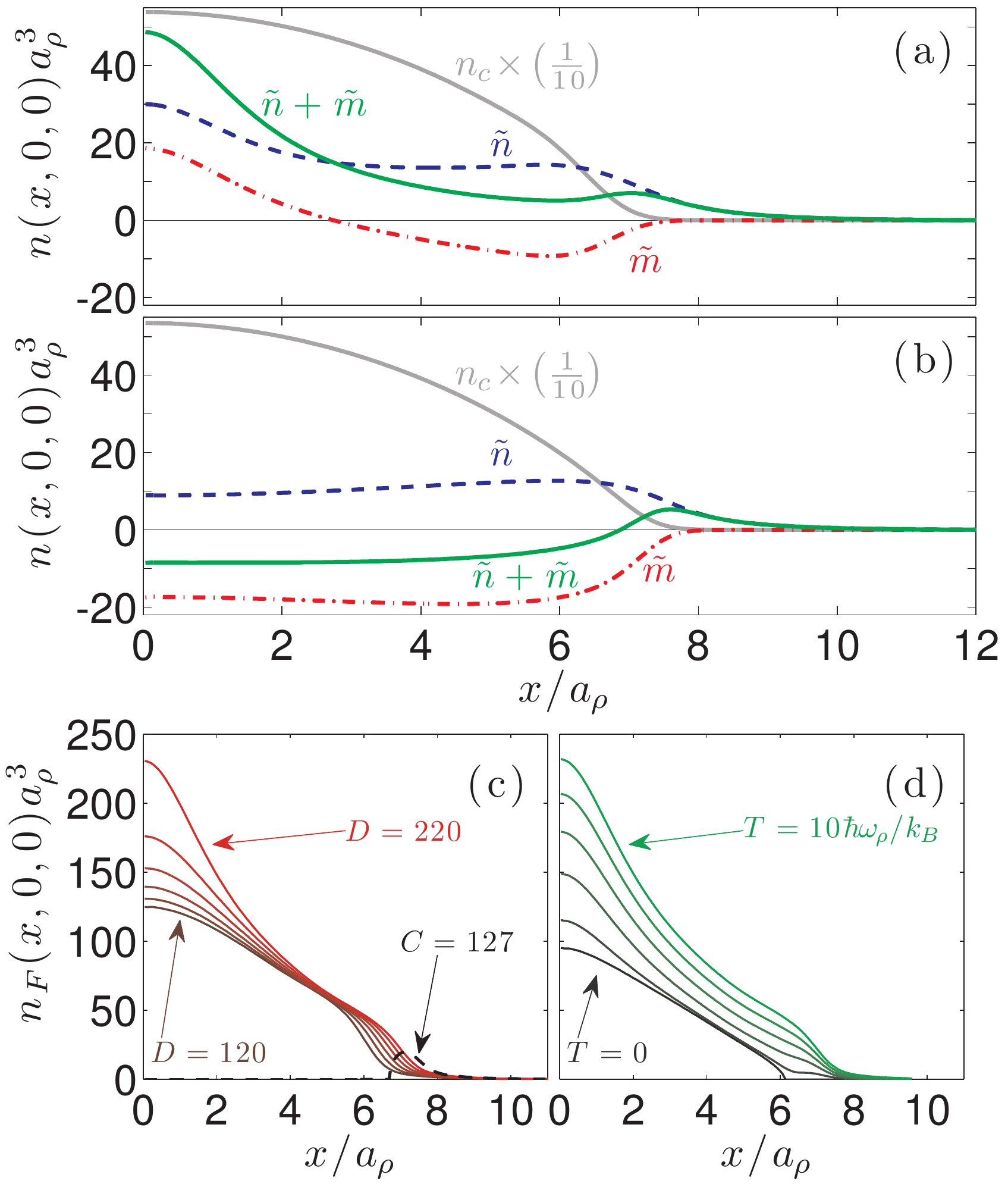}
\caption{(color online) Comparison between densities of (a)  a purely dipolar condensate with $D=220$ and (b) a purely contact interaction condensate with $C=127$.  Figures show the (blue dashed) normal density $\tilde{n}$, (red dash dot) anomalous density $\tilde{m}$,   (green solid) $\tilde{n}+\tilde{m}$, and (grey solid) the scaled condensate density, all evaluated along the $x$ axis. 
(c) Shows the development of the fluctuation density $n_F$ (see text) with dipolar strength for a purely dipolar condensate  ($D=120$ to $D=220$  in steps of $\Delta D=20$). The dashed line shows the real part of $n_F$ for the purely contact case of $C=127$. These results are at $T=10\hbar\omega_{\rho}/k_B$.
 (d) Shows the development of the fluctuation density (see text) with temperature for the purely dipolar condensate with $D=220$ ($T=0$ to $T=10\hbar\omega_{\rho}/k_B$ in steps of $\Delta T=2\hbar\omega_{\rho}/k_B$). Other parameters $\lambda=20$ and  $N_0=25\times10^3$.  
 \label{Fig:posden2}}
\end{center}
\end{figure}

The anomalous density \cite{BlaizotRipka}, 
which characterizes pairing correlations in the thermal component of the field, is given by
\begin{equation}
\tilde{m}(\mathbf{x})\equiv\langle\hat{\delta}\hat{\delta}\rangle=-\sum_j(2\bar{n}_j+1)u_j(\mathbf{x}) v_j^*(\mathbf{x}).\label{mtilde}
\end{equation} 
Because pairing is strongly influenced by the nature of the interactions (i.e.~attractive versus repulsive interactions) it is useful to compare the anomalous density between dipolar  and contact  condensates which we do in Fig.~\ref{Fig:posden2}(a) and (b).
The parameters of this comparison are the same as those used in Figs.~\ref{Fig:posden}(a1) and (b1). For the contact case [Fig.~\ref{Fig:posden2}(b)]   the anomalous density exhibits behavior seen in other studies of a trapped condensate with repulsive contact interactions (see \cite{Wright2011a} and references therein); notably $\tilde{m}$ is negative in the region where the condensate density is significant.  For the dipolar condensate [Fig.~\ref{Fig:posden2}(a)] the anomalous density is similar to the contact case near the condensate surface ($|x|\gtrsim4a_\rho$), but has strikingly different behavior near the trap center, where it turns positive and forms a peak.

The anomalous density manifests itself in the density  fluctuations of the system. These can be characterized by the second order correlation function \cite{Naraschewski1999a}
\begin{align}
G^{(2)}(\mathbf{x})&=\langle \hat{\Psi}^\dagger(\mathbf{x})\hat{\Psi}^\dagger(\mathbf{x})\hat{\Psi} (\mathbf{x})\hat{\Psi} (\mathbf{x})\rangle,\\
&=n_c^2+4n_c\tilde{n}+2n_c\tilde{m}+2\tilde{n}^2+\tilde{m}^2,\label{G2Bog}\\
&\approx n^2\left[1+\frac{2(\tilde{n}+\tilde{m})}{n_c}\right],\label{G2normA}
\end{align}
where $n_c=|\psi_0|^2$ is the condensate density and $n=n_c+\tilde{n}$ is the total density. Result (\ref{G2Bog}) is obtained using Wick's theorem \cite{BlaizotRipka,Holzmann1999a} (also see \footnote{The non-local (two-point)  $G^{(2)}$ correlation function was calculated for a trapped dipolar condensate in Ref.~\cite{Ticknor2012b}. Those calculations were performed within a quasi-2D approximation at $T=0.5T_c$, but neglected the contribution of $\tilde{m}$.} and \cite{Baillie2012a}). Expression (\ref{G2normA}) is accurate to first order in $\tilde{n}/n_c$ and $\tilde{m}/n_c$   and  shows how the density fluctuations are sensitive to the sum of the non-condensate and anomalous densities. We also show $\tilde{n}+\tilde{m}$  in Figs.~\ref{Fig:posden2}(a) and (b). For the contact gas this quantity  is less than zero, which shows the suppression of density fluctuations, except near the surface where $\tilde{n}\gtrsim n_c$. In contrast, for the dipolar condensate $\tilde{n}+\tilde{m}$ is positive  and peaked at the trap center. 

To put the relative scale of the density fluctuations into context, it is convenient to define the \textit{fluctuation density} as the root-mean-square of the density fluctuations, i.e.~
 \begin{align}
 n_F(\mathbf{x})&= \sqrt{G^{(2)}(\mathbf{x})-n(\mathbf{x})^2}.
 \end{align}
In Figs.~\ref{Fig:posden2}(c) and (d) we examine $n_F(\mathbf{x})$ for the dipolar gas as a function of the DDI parameter and temperature.   In Fig.~\ref{Fig:posden2}(c) the total density at the trap center decreases with increasing $D$; i.e.~the total density decreases from $n(\mathbf{0})=783\,a_{\rho}^{-3}$ at $D=120$, to  $n(\mathbf{0})=568\,a_{\rho}^{-3}$ at $D=220$. Over this same range of $D$ values the central fluctuation density increases from $n_F(\mathbf{0})=125\,a_{\rho}^{-3}$  to $n_F(\mathbf{0})=231\,a_{\rho}^{-3}$. 
For the purposes of comparison, in Fig.~\ref{Fig:posden2}(c) we also show the real part \footnote{For the contact case $G^{(2)}(\mathbf{x})<n(\mathbf{x})^{2}$ in the central region  and the fluctuations are suppressed below the Poissonian level.}   of $n_F$ for the contact case.  In this case the fluctuation density is only non-zero near the  surface of the condensate.  To leading order in $\tilde{n}$ and $\tilde{m}$ the fluctuation density is given by $n_F\approx\sqrt{2n_c(\tilde{n}+\tilde{m})}$.

\subsection{Interplay of contact and dipole interactions}
 \begin{figure}[!tbh]
\begin{center}
\includegraphics[width=3.4in]{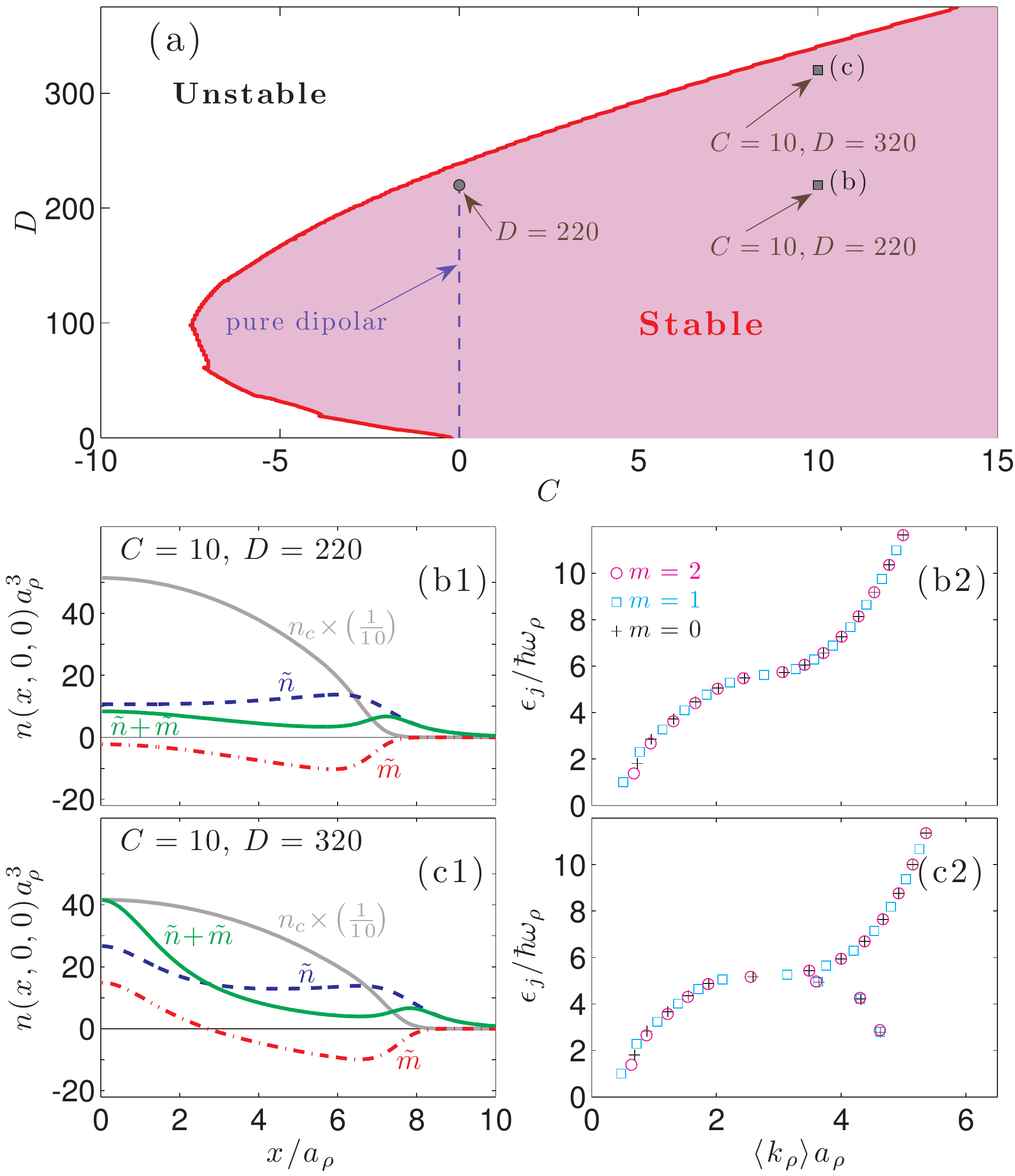}
\caption{(color online) (a) Stability phase diagram in interaction parameter space for atoms in a $\lambda=20$ trap. The shaded region indicates where a stable solution exists. Dashed line indicates the case of purely dipolar interactions ($C=0$) which were primarily considered earlier in the paper, with the particular case of $C=0,D=220$ indicates by a filled circle.
The squares indicate two sets of parameters with non-zero contact interaction that we examine. The square labeled (b) is  $C=10, D=220$ with results  given in (b1) and (b2). The square labeled  (c) is  $C=10, D=320$ with results  given in (c1) and (c2).
(b1,c1)  the (blue dashed) normal density $\tilde{n}$, (red dash dot) anomalous density $\tilde{m}$,   (green solid) $\tilde{n}+\tilde{m}$, and (grey solid) the scaled condensate density, all evaluated along the $x$ axis.  (b2,c2) The mapped dispersions. Other parameters $\lambda=20$,  $N_0=25\times10^3$, and  $T=10\hbar\omega_{\rho}/k_B$.  
 \label{Fig:contdip}}
\end{center}
\end{figure}

In general a dipolar condensate will have non-zero contact and dipolar interactions. Here we briefly consider how contact interactions modify our predictions. In Fig.~\ref{Fig:contdip}(b1) we show the non-condensate and anomalous density for a dipolar condensate with $C=10$ and $D=220$. These results compared to those in Fig.~\ref{Fig:posden2}(a), which were for the purely dipolar case of $D=220$, reveal that the added contact interaction causes the roton density peak to be suppressed, and the anomalous density to be everywhere negative. Examining the mapped dispersion relation [Fig.~\ref{Fig:contdip}(b2)] we see that for this case the spectrum no longer has a roton minimum [c.f.~Fig.~\ref{Fig:posden}(a2) for the purely dipolar case]. This occurs because increasing the contact interaction stabilizes the system, i.e.~moves the system further from the stability boundary. To illustrate this we have shown the stability phase diagram for the $\lambda=20$ trap in  Fig.~\ref{Fig:contdip}(a). By increasing the dipole strength and approaching the stability boundary, the roton re-emerges in the spectrum and the density peak becomes apparent at the trap center [see Figs.~\ref{Fig:contdip}(c1), (c2)].  We note that in experiments it is usually convenient to tune the contact interaction via a Feshbach resonance  to explore the stability of dipolar condensates (e.g.,~see \cite{Koch2008a}).

\subsection{Momentum space density and depletion}
 \begin{figure}[tbh]
\begin{center}
\includegraphics[width=3.25in]{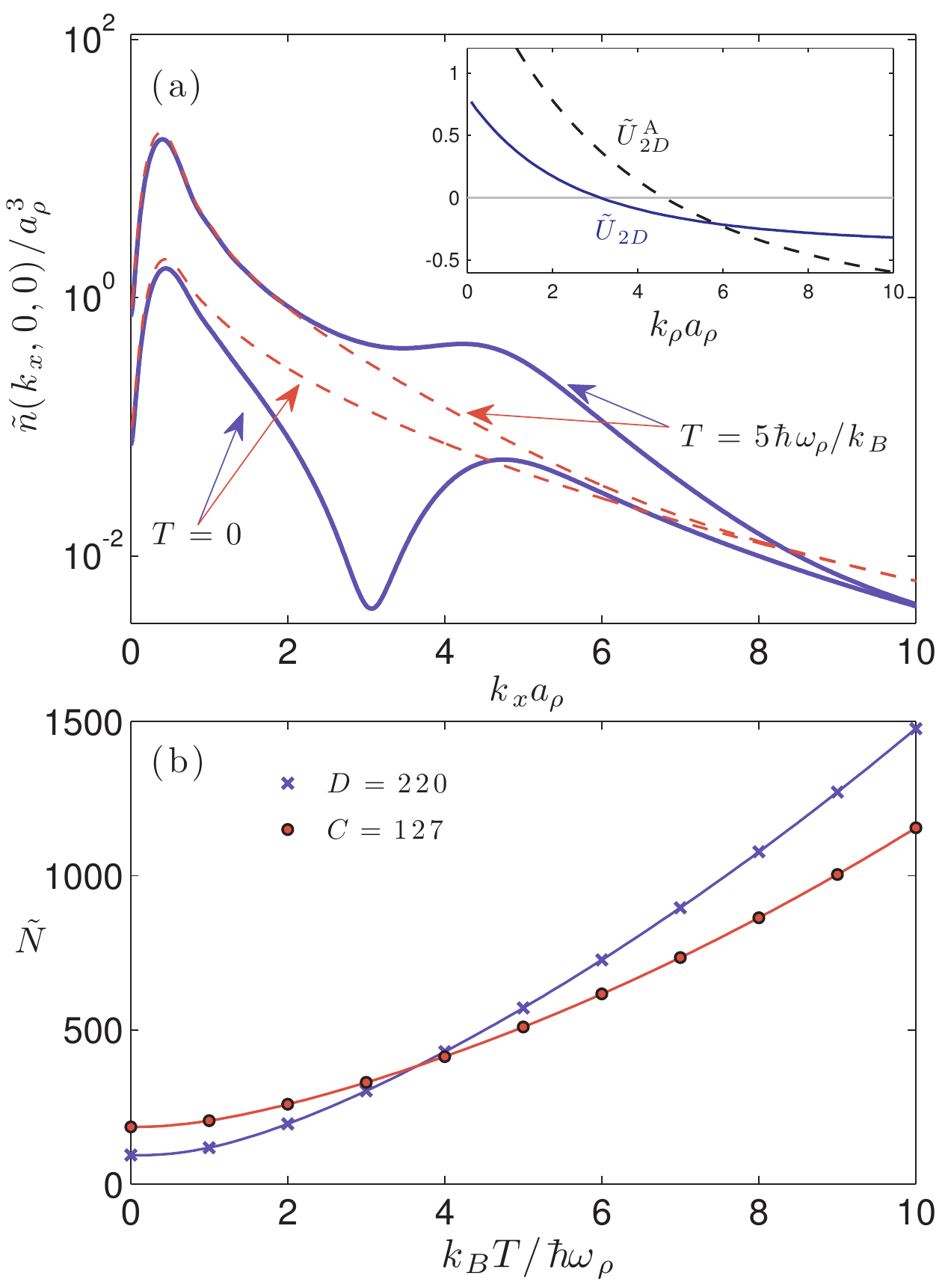} 
\caption{(color online)  (a) Non-condensate density in momentum space  for a dipolar condensate with $D=220$ ($C=0$) (blue lines) and a  contact interaction condensate with $C=127$ ($D=0$) (red dashed line). Inset: effective 2D interaction obtained using the condensate profile $\tilde{U}_{2D}$ and analytically using a Gaussian profile $\tilde{U}_{2D}^{\mathrm{A}}$ (see Appendix for details).  (b) Total depletion as a function of the temperature for the contact and dipolar systems considered in (a).  Other parameters as in Fig.~\ref{Fig:posden}.
 \label{Fig:mtm}}
\end{center}
\end{figure} 

Because the rotons develop in a specific momentum range [see Fig.~\ref{Fig:posden}(a2)] it is interesting to consider how they manifest in the momentum space density of the non-condensate. Indeed,   Mazets \textit{et al.}~\cite{Mazets2004a} considered the momentum distribution in a quasi-1D (cigar shaped) system with laser induced  DDIs. Their results  showed that the presence of a roton minimum in the excitation spectrum was revealed by a local peak in the non-condensate momentum distribution at a momentum corresponding to the roton. Similarly,  Jona-Lasinio \textit{et al.}~\cite{JonaLasinio2013} have shown using time dependent GPE simulations that a halo in momentum space can develop for a pancake dipolar BEC (e.g.~by quenching interactions).

Here we examine the non-condensate density in momentum space, which is given by
\begin{equation}
\tilde{n}(\mathbf{k})=\sum_j\left[\bar{n}_j|\bar u_j(\mathbf{k})|^2+(\bar{n}_j\!+\!1)|\bar v_j(\mathbf{k})|^2\right].\label{ntildek}
\end{equation}
In Fig.~\ref{Fig:mtm}(a) we compare  $\tilde{n}(\mathbf{k})$ for dipolar and contact cases at $T=0$ and $T=5\,\hbar\omega_{\rho}/k_B$.

The $T=0$ results reveal  the momentum distribution of the quantum depletion (i.e.~atoms excited out of the condensate due to interactions). A feature of  note is the strong suppression of $\tilde{n}(\mathbf{k})$ for the dipolar case for momenta near $k_x\sim3/a_\rho$.  This can be understood by considering the momentum dependence of the interaction [see inset to Fig.~\ref{Fig:mtm}(a) and Appendix]. This shows that the interaction crosses over from being positive (repulsive) for $k_{\rho}\lesssim3/a_{\rho}$ to attractive at $k_{\rho}\gtrsim3/a_{\rho}$. Thus for  $k_{\rho}\sim3/a_\rho$ the interaction is effectively zero, explaining why few atoms are depleted at this momentum. The contact interaction gas does not have this suppression  and as a result at $T=0$ the total depletion, $\tilde{N}=\int d\mathbf{x}\,\tilde{n}(\mathbf{x})$,  is larger [see Fig.~\ref{Fig:mtm}(b)]. The local maximum that occurs at $k_x\sim5/a_{\rho}$ \footnote{We note that the rotons emerge with a wavevector set by the $z$ confinement length, i.e.~$k\sim 1/a_z$. } is associated with the same roton modes that gave rise to the  roton peak in the position space non-condensate density (also see \cite{JonaLasinio2013}).  

As temperature increases the softer spectrum of the dipolar condensate becomes thermally activated more rapidly than the contact case, particularly in the vicinity of the roton [e.g.~see the $T=5\hbar\omega_{\rho}/k_B$ result in Fig.~\ref{Fig:mtm}(a)].  Indeed, at all temperatures we have considered (up to $10\,\hbar\omega_{\rho}/k_B$) we observe a local maximum in the momentum density at the roton wavevector.

\section{Conclusion and outlook} 
In this paper we have explored the  behavior of key quantities for a trapped dipolar gas: the non-condensate density, in both position and momentum space, and the anomalous density. To our knowledge this is the first time these quantities have been reported for this system and compared to the better understood case of a condensate with contact interactions.
 The concept of a roton gas forming in the central region of a pancake dipolar condensate was recently proposed in Ref.~\cite{JonaLasinio2013} based on a  local density analysis in a quasi-2D approximation. Our results support this concept with general calculations that reveal an intrinsic 3D character to the rotons, and demonstrate the effect of temperature. Importantly we show that the roton gas emerges as a peak in the non-condensate at trap center, and that in this region the anomalous density is positive leading to large density fluctuations. In momentum space the non-condensate density reveals the presence of rotons as a halo with radius set by the roton momentum. 

Here we have restricted our results to  low temperatures because we have not included the  back-action of the non-condensate on itself or the condensate. For our results with $D=220$ and $T=10\hbar\omega_\rho/k_B$ ($\sim0.13T_c$) the density fluctuations are observed to be significant, suggesting that this back-action may already be important.  We note that several recent finite temperature studies have been performed using Hartree-Fock \cite{Bisset2012} and Hartree-Fock Bogoliubov \cite{Ronen2007b,Ticknor2012a} approaches in which the non-condensate is obtained self-consistently with the condensate. However, all of these calculations have been made employing a Popov approximation \cite{Griffin1996a} in which the anomalous density is  neglected. Our results show that the anomalous density  significantly enhances density fluctuations, putting into question the appropriateness of the Popov approximation for the dipolar gas. We also note a recent discussion of the validity of Bogoliubov theory for the uniform dipolar system \cite{Boudjemaa2013a}. Alternatively, a promising direction for the development of a suitable theory for warm dipolar BECs may be provided by classical field approaches \cite{Blakie2009e,*cfieldRev2008,Pawlowski2013}. These approaches naturally include the anomalous density (e.g.~see \cite{Wright2011a}) and should be valid when density fluctuations are large. 

 Rotons have yet to be observed in experiments with dipolar condensates and there remains great interest in developing robust signatures to aid in identifying them. Our results show that density fluctuations provide a clear signature of the presence of rotons. We emphasize that experimental measurements of density fluctuations are always performed using finite size cells (e.g.~due to optical resolution limits), and the length scale of the cells plays a fundamental role in such measurements, as discussed in Ref.~\cite{Bisset2013a}.

\section*{Acknowledgments}  This work was supported by the Marsden Fund of New Zealand (contracts UOO0924 and UOO1220).

\appendix
\section{Effective 2D interaction in $\mathbf{k}$-space}
We obtain the 2D (in-plane) DDI in $k$-space as
\begin{equation}
\!\tilde{U}_{\mathrm{2D}}({ {k}_\rho}) =  \int \frac{dk_z}{2\pi}\,\tilde{U}_{\mathrm{dd}}({ {k}_\rho},k_z)\tilde{\lambda}_0 (k_z)^2,\label{V2DGPE}
\end{equation} where ${ {k}_\rho}=\sqrt{k_x^2+k_y^2}$, $\tilde\lambda_0(k_z)$ is the Fourier transform of the normalized $z$-profile of the condensate at $x=y=0$, i.e.~
\begin{equation}
\lambda_0(z)=\frac{|\psi_0(\mathbf{0},z)|^2}{\int dz'|\psi_0(\mathbf{0},z')|^2}
\end{equation}
 and 
 \begin{equation}
 \tilde{U}_{\mathrm{dd}}( {k}_{\rho},k_z)=g_{\mathrm{dd}}\left(3\frac{k_z^2}{k_{\rho}^2}-1\right),
 \end{equation}
  is the Fourier transform of $U_{\mathrm{dd}}$.  
Note that in the limit of vanishing interactions, where the $z$ shape of the condensate is a Gaussian  (independent of $\rho$), $\tilde{U}_{\mathrm{2D}} ({{k}_\rho})$ reduces to the analytic result
\begin{equation} 
\tilde{U}^{\mathrm{A}}_{\mathrm{2D}}({{k}_\rho})= \frac{g_{\mathrm{dd}}}{\sqrt{2\pi}a_z} F_{\perp}\left(\tfrac{1}{\sqrt{2}}k_\rho a_z \right),\label{V2D}
\end{equation}
where $F_{\perp}(Q)=2-3\sqrt{\pi}Qe^{Q^2}\mathrm{erfc}\left(Q\right)$,  with $a_z=\sqrt{\hbar/m\omega_z}$ the $z$ confinement length.  Our results in the inset to Fig.~\ref{Fig:mtm}(a), show that $\tilde{U}^{\mathrm{A}}_{\mathrm{2D}}$ is a poor approximation for $D=220$, demonstrating that the quasi-2D approximation is invalid in this regime.


%

\end{document}